\documentclass[prd,twocolumn,twoside,preprintnumbers,superscriptaddress,nofootinbib]{revtex4-1}
\usepackage[a4paper, hdivide={1.91cm,,1.165cm}, vdivide={1.83cm,,3.2cm}]{geometry}

\usepackage[utf8]{inputenc}
\usepackage[english]{babel}
\usepackage{amsmath}
\usepackage{amssymb}
\usepackage{graphicx}
\usepackage{bbold}
\usepackage{dcolumn}
\usepackage[hyperfootnotes=false]{hyperref}
\usepackage{xspace}

\usepackage{colortbl}
\definecolor{Gray}{gray}{0.95}
\definecolor{RGray}{gray}{0.85}
\definecolor{CGray}{gray}{0.92}
\makeatletter

    \def\CT@@do@color{%
      \global\let\CT@do@color\relax
            \@tempdima\wd\z@
            \advance\@tempdima\@tempdimb
            \advance\@tempdima\@tempdimc
    \advance\@tempdimb\tabcolsep
    \advance\@tempdimc\tabcolsep
    \advance\@tempdima2\tabcolsep
            \kern-\@tempdimb
            \leaders\vrule
                    \hskip\@tempdima\@plus  1fill
            \kern-\@tempdimc
            \hskip-\wd\z@ \@plus -1fill }
\makeatother

\newcommand{\bea}{\begin{eqnarray}}
\newcommand{\eea}{\end{eqnarray}}

\usepackage[dvipsnames]{xcolor}

\allowdisplaybreaks

\begin{document}
\preprint{\vbox{\hbox{PSI-PR-16-13}}, \vbox{\hbox{IFIC/16-78}}, \vbox{\hbox{ZU-TH-40-16}}}

\title{ {Lepton Flavor Non-Universality in B decays from Dynamical Yukawas}}

\author{ Andreas Crivellin}
\email{andreas.crivellin@cern.ch}
\affiliation{Paul Scherrer Institut, CH--5232 Villigen PSI, Switzerland}
\author{Javier Fuentes-Mart\'in}
\email{javier.fuentes@ific.uv.es}
\affiliation{Instituto de F\'isica Corpuscular, Universitat de Val\`encia - CSIC, E-46071 Val\`encia, Spain}
\author{Admir Greljo}
\email{admir@physik.uzh.ch}
\affiliation{Physik-Institut, Universit\"at Z\"urich, CH-8057 Z\"urich, Switzerland}
\affiliation{Faculty of Science, University of Sarajevo, Zmaja od Bosne 33-35, \\ 71000 Sarajevo, Bosnia and Herzegovina}
\author{Gino Isidori}
\email{isidori@physik.uzh.ch}
\affiliation{Physik-Institut, Universit\"at Z\"urich, CH-8057 Z\"urich, Switzerland}

\begin{abstract}
The basic features of quark and lepton mass matrices can be successfully explained by natural minima of a generic potential with dynamical Yukawa fields invariant under the $[\mathrm{SU(3)}]^5\times \mathcal{O}(3)$ flavor symmetry. If this symmetry is gauged, in order to avoid potentially dangerous Goldstone bosons, and small perturbations are added to exactly fit the observed pattern of fermion masses, the spectrum of massive flavor gauge bosons can naturally explain the hints for new physics in $b\to s \ell^+\ell^-$ transitions, including $R_K$. In particular, the desired pattern of the Standard Model Yukawa couplings is compatible with a gauged $\mathrm{U(1)}_q$ in the quark sector, and $\mathrm{U(1)}_{\mu-\tau}$ in the lepton sector spontaneously broken around the TeV scale. In order to explain the aforementioned experimental hints, the corresponding neutral gauge bosons are required to mix, yielding to potentially observable signals in dimuon resonance searches at the LHC.
\end{abstract}

\maketitle

\section{Introduction}
\label{sec:intro}

With the Higgs boson discovery at the LHC, the Standard Model (SM) of particle physics stands out as a great success story. Since no new particles were observed so far, this only amplifies the importance of some long-standing questions. In this respect, the SM flavor puzzle --the lack of explanation for the peculiar structure of fermion masses and mixing angles-- with masses spanning several orders of magnitude and very hierarchical mixing in the quark sector in contrast to the anarchical structure in the lepton sector, is still a critical point any theory of physics beyond the SM should address.

While most flavor observables agree very well with the SM, there are significant deviations from the theory prediction in $b\to s\mu^+\mu^-$ processes (see for example recent review~\cite{Crivellin:2016ekz}). If confirmed, these could be a guideline to uncover the flavor structure of physics beyond the SM. In more details, the current theoretical and experimental situation is the following: LHCb~\cite{Aaij:2014ora} measured the ratio
\begin{equation}
R_K=\dfrac{{\rm Br}[B\to K \mu^+\mu^-]}{{\rm Br}[B\to K e^+e^-]}=0.745^{+0.090}_{-0.074}\pm 0.036\,,
\end{equation}
which deviates from the theoretically clean SM prediction  {$R_K^{\rm SM}=1$ up to $\mathcal{O}(1\%)$~\cite{Bordone:2016gaq}} by $2.6\,\sigma$. In addition, LHCb has reported deviations from the SM predictions~\cite{Descotes-Genon:2014uoa,Descotes-Genon:2015xqa,Altmannshofer:2014rta,Straub:2015ica,Hurth:2016fbr} in the decay $B\to K^*\mu^+\mu^-$ (mainly in the angular observable $P_5^\prime$~\cite{DescotesGenon:2012zf}) with a significance of about $3\,\sigma$~\cite{Aaij:2013qta,LHCb:2015dla}. Very recently, a (less precise) Belle measurement~\cite{Abdesselam:2016llu} confirmed the so-called $P_5^\prime$ anomaly at the $2\sigma$ level. Moreover, the measurement of ${\rm Br}[B_s\to\phi\mu^+\mu^-]$ disagrees with the SM prediction~\cite{Horgan:2013pva, Horgan:2015vla} by about $3\,\sigma$~\cite{Altmannshofer:2014rta}.

Interestingly enough, all these discrepancies can be explained in a model-independent approach by a rather large new-physics (NP) contribution (at the level of $- 25\%$ of the SM prediction) to effective Flavor Changing Neutral Current (FCNC)  operators involving the product of $\bar s b$ and $\bar \mu \mu$ currents (whose Wilson coefficients are denoted $C_{9,10}^{(\prime)\mu\mu}$, see Sect.~\ref{sec:Pheno}). Three symmetry based solutions (taking into account at most two non-zero Wilson coefficients) give a very good fit to data: NP in $C_9^{\mu\mu}$ only~\cite{Descotes-Genon:2013wba,Altmannshofer:2013foa,Alonso:2014csa,Hiller:2014yaa,Ghosh:2014awa,Altmannshofer:2015sma,Descotes-Genon:2015uva,Hurth:2016fbr}, 
$C_9^{\mu\mu}|_{\rm NP} =-C_{10}^{\mu\mu}|_{\rm NP} $~\cite{Ghosh:2014awa,Altmannshofer:2015sma,Descotes-Genon:2015uva,Hurth:2016fbr} and $C_9^{\mu\mu}|_{\rm NP} =-C_{9}^{\prime\mu\mu}|_{\rm NP} $~\cite{Descotes-Genon:2015uva}. It is encouraging that the value for $C^{\mu\mu}_{9,10}|_{\rm NP} $ required to explain $R_K$ (with $C^{ee}_9|_{\rm NP} =0$) is of the same order as the one needed for $B\to K^*\mu^+\mu^-$ and $B_s\to\phi\mu^+\mu^-$ and such scenarios are preferred over the SM by $4$--$5\sigma$. 

Many models proposed to explain the $b\to s \mu^+\mu^-$ data contain a massive neutral gauge boson ($Z^\prime$) which generates a tree-level contribution to $C_9^{\mu\mu}$~\cite{Descotes-Genon:2013wba,Gauld:2013qba,Buras:2013qja,Gauld:2013qja,Buras:2013dea,Altmannshofer:2014cfa,Crivellin:2015mga,Crivellin:2015lwa,Niehoff:2015bfa,Crivellin:2015era,Sierra:2015fma,Belanger:2015nma,Celis:2015ara,Allanach:2015gkd,Altmannshofer:2014cfa,Altmannshofer:2015mqa,Kim:2016bdu,Celis:2016ayl,Boucenna:2016qad,Megias:2016bde}, and if the $Z^\prime$ couples differently to muons and electrons, $R_K$ can be explained simultaneously.  Alternatively, models with weak triplet vectors~\cite{Greljo:2015mma,Boucenna:2016wpr,Boucenna:2016qad}, leptoquarks~\cite{Gripaios:2014tna,Becirevic:2015asa,Varzielas:2015iva,Alonso:2015sja,Calibbi:2015kma,Barbieri:2015yvd,Bauer:2015knc,Becirevic:2016yqi,Becirevic:2016oho,Sahoo:2016pet,Bhattacharya:2016mcc}, or heavy new scalars and fermions~\cite{Gripaios:2015gra,Arnan:2016cpy} were also proposed.

Coming back to the SM flavor puzzle, an interesting approach to reproduce the main features of quark and lepton mass matrices is to assume the Yukawa couplings are generated from dynamical fields, whose background values minimize a generic potential invariant under a large non-Abelian flavor symmetry group, such as  $[\mathrm{SU(3)}]^5\times \mathcal{O}(3)$~\cite{D'Ambrosio:2002ex,Cirigliano:2005ck}. Employing general group theory arguments, the natural extrema of such potential, corresponding to maximally unbroken subgroups, robustly predict large (zero) third (first two) generation quark mass, and a trivial CKM matrix (equal to the identity matrix). In the lepton sector, the same approach leads to a solution with  hierarchical charged lepton masses accompanied by at least two degenerate Majorana neutrinos, with potentially large $\theta_{12}$ mixing angle, $\theta_{23}=\pi/4$, $\theta_{13}=0$ and one maximal Majorana phase~\cite{Alonso:2013nca}. Adding small perturbations to this picture, quantitatively reconciles well with observations.

In such framework, in order to avoid massless Goldstone bosons, it is natural to expect that the flavor symmetry is gauged (see e.g.~\cite{Grinstein:2010ve,Albrecht:2010xh,Alonso:2016onw}). 
The complete spectrum of the corresponding massive vector bosons is quite complicated and may span several orders of magnitude. However, a large fraction of such states may be irrelevant at low energies. In this paper we focus on the phenomenology of the potentially lightest vector states, associated to some of the residual unbroken subgroups. In particular, this setup naturally leads to a gauged $L_\mu-L_\tau$ symmetry~\cite{He:1990pn,Foot:1990mn,He:1991qd,Binetruy:1996cs,Bell:2000vh,Choubey:2004hn,Dutta:1994dx,Heeck:2011wj,Heeck:2014qea} in the lepton sector~\cite{Alonso:2013nca} and an independent Abelian symmetry in the quark sector, $\mathrm{U(1)}_q$, that may have interesting implications for the observed deviations from the SM in $b\to s\mu^+\mu^-$. Interestingly, gauging a linear combination of the two $\mathrm{U(1)}$'s was already proposed in Ref.~\cite{Crivellin:2015lwa} as a solution of the LHCb anomalies. While this was purely phenomenologically motivated, here we suggest  that such symmetries might be connected with the observed pattern of fermion masses and mixings. In contrast to Ref.~\cite{Crivellin:2015lwa}, the presence of two neutral gauge bosons make the collider signatures of this model quite different, allowing for lighter gauge bosons.

\begin{figure}[t]
\hspace{-10pt}\includegraphics[width=1.0\hsize]{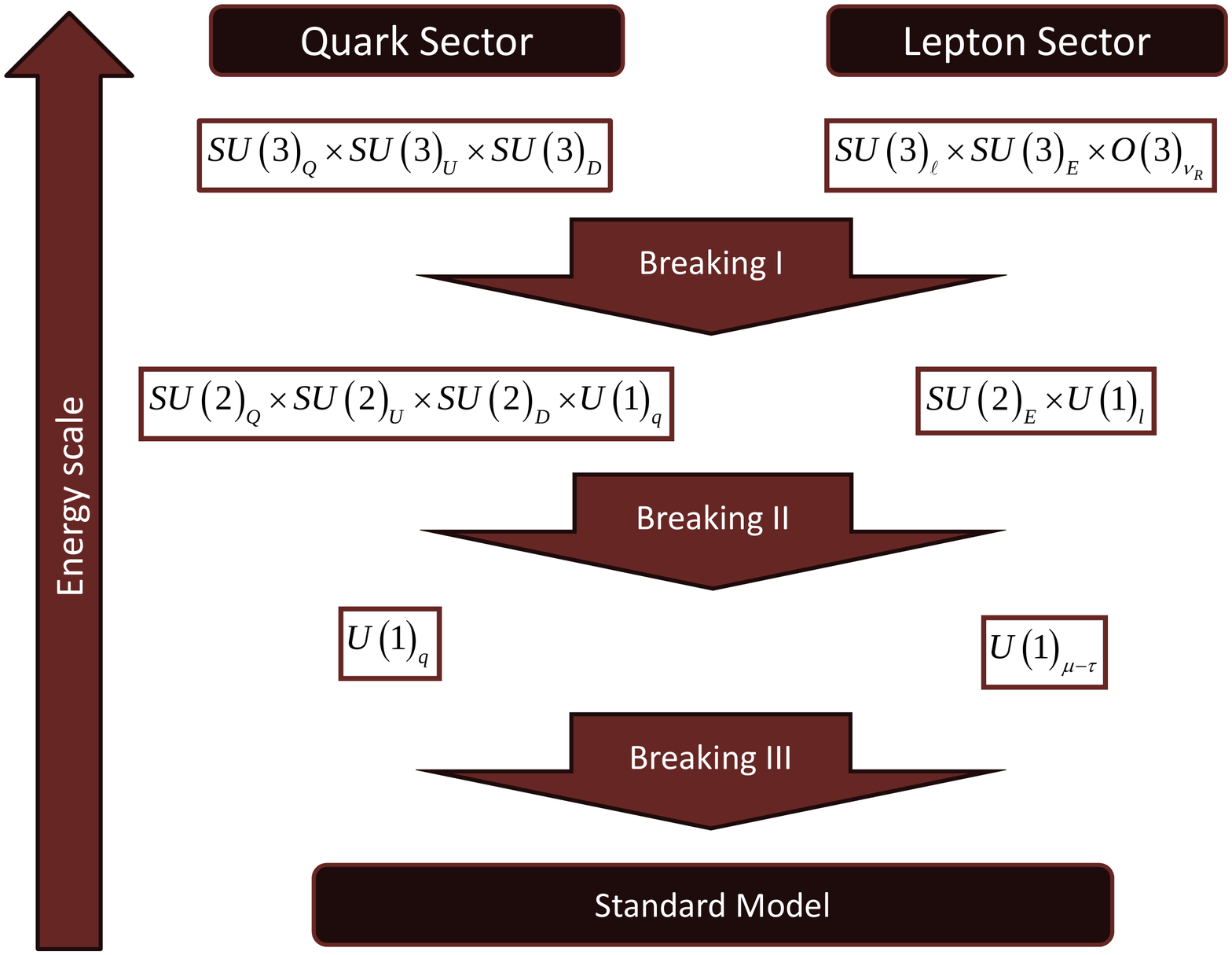}
\caption{\small \sf Illustration of the symmetry breaking pattern in the quark and lepton sectors. 
\label{fig:SB} }
\end{figure}

\section{Natural minima perturbations and lowest-lying \texorpdfstring{$Z'$}{Z'} bosons}
\label{sec:scenario}

Building on Ref.~\cite{Alonso:2013nca}, we consider the extension of the SM gauge sector by the maximal flavor symmetry in the limit of vanishing Yukawa couplings and non-vanishing Majorana mass term for the right-handed neutrinos: $\mathcal{G}=\mathrm{SU(3)}_Q \times \mathrm{SU(3)}_D \times \mathrm{SU(3)}_U \times \mathrm{SU(3)}_\ell \times \mathrm{SU(3)}_E\times\mathcal{O}(3)_{\nu_R}$. Here, $Q$ ($\ell$) corresponds to the left-handed quarks (leptons) while $U$ ($D$) and $E$ stand for the right-handed up (down) quarks and charged leptons, respectively. Neutrino masses are accounted for via the see-saw mechanism by introducing three right-handed neutrinos. Natural extrema of a generic Yukawa scalar potential break the flavor symmetry to a maximal subgroup, providing a good first-order explanation of the fermion masses and mixings, both in quark and lepton sectors~\cite{Alonso:2013nca}. In a second step, perturbations are introduced (e.g. via extra scalar fields~{\cite{Alonso:2011yg,Fong:2013dnk}}), breaking the residual flavor symmetry and fitting the observed masses and mixing angles. Below we present a  detailed discussion of the sequential breaking patterns for both sectors, identifying the appearance of the two residual $\mathrm{U(1)}$ symmetries. Finally, the spectrum and the couplings of the corresponding $Z^\prime$ bosons are discussed. The global picture is illustrated in Fig.~\ref{fig:SB}.

\subsection{Quark sector}
\label{subsec:quark}

{\bf Step I:} Following the approach in Ref.~\cite{Alonso:2013nca}, we choose the hierarchical natural minima of the flavon fields in such a way that the following form for the Yukawa couplings is generated:
\begin{align}\label{eq:nat_min_q}
Y_u^{(0)}=\mathrm{diag}\left(0,0,y_t\right)\,,\qquad Y_d^{(0)}=\mathrm{diag}\left(0,0,y_b\right)\,,
\end{align}
leading to a diagonal CKM matrix and resulting in the following flavor symmetry breaking pattern
\begin{align}
\begin{aligned}
&\mathrm{SU(3)}_Q\times\mathrm{SU(3)}_U\times\mathrm{SU(3)}_D\to\\
&\hspace{1.6cm}\mathrm{SU(2)}_Q\times\mathrm{SU(2)}_U\times\mathrm{SU(2)}_D\times\mathrm{U(1)}_q\,.
\end{aligned}
\end{align}
This solution provides a good starting point for the explanation of fermion masses and mixings. Here $\mathrm{U(1)}_q$ is defined by the subgroup of the flavor symmetry given by
\begin{align}
\mathrm{U(1)}_q:\; \exp(i \alpha \lambda_8^Q)\otimes\exp(i \alpha \lambda_8^U)\otimes\exp(i \alpha \lambda_8^D)\,,
\end{align}
with $\lambda_8^I$ denoting the Gell-Mann matrix, which corresponds to the following charge assignments in generation space
\begin{align}
X_Q=X_U=X_D= \left(-\frac{1}{2},-\frac{1}{2},1\right)\,.
\end{align}

\smallskip
{\bf Step II:}  Perturbations around this minima are needed to provide the correct description of the observed masses and mixing angles. However, since the natural minima are very stable, it is not possible to induce these perturbations through one-loop corrections and (or) higher dimensional operators including Yukawa fields only~\cite{Espinosa:2012uu}. In Ref.~{\cite{Alonso:2011yg,Espinosa:2012uu,Fong:2013dnk}} it was shown that the correct perturbations can be introduced in a natural way through the inclusion of additional scalars in reducible representations of the flavor group. In this letter, we focus on the perturbations yielding the following flavor symmetry breaking pattern
\begin{align}\label{eq:SSB_patt}
\mathrm{SU(2)}_Q\times\mathrm{SU(2)}_U\times\mathrm{SU(2)}_D\times\mathrm{U(1)}_q\to\mathrm{U(1)}_q\,,
\end{align}
at a very high scale, followed by a subsequent breaking of $\mathrm{U(1)}_q$ around the TeV scale ({\bf Step III}). 

Finally, the perturbations to Eq.~\eqref{eq:nat_min_q} take the form
\begin{align}\label{eq:final_Yuk}
Y_{u(d)}^{(1)}=
\begin{pmatrix}
\epsilon_{u(d)}^{11} & \epsilon_{u(d)}^{12} &  \delta^1_{u(d)}\\
\epsilon_{u(d)}^{21} & \epsilon_{u(d)}^{22} & \delta^2_{u(d)}\\
\sigma_{u(d)}^1 & \sigma_{u(d)}^2 & 0
\end{pmatrix}
\,,
\end{align}
and the full Yukawa matrix is given by $Y_{u(d)}=Y^{(0)}_{u(d)}+Y_{u(d)}^{(1)}$. The correct masses and mixings are obtained for
\begin{equation}
|\epsilon_{u(d)}^{ij}| \ll |\delta_{u(d)}^{i}| \ll 1~.
\end{equation}
For simplicity, we assume negligible $\sigma_{u(d)}^i$: these terms are not required for fitting the CKM matrix and their absence/smallness 
is welcome to suppress right-handed FCNCs. In a given model,   $\epsilon_{u(d)}$ and $\delta_{u(d)}$ are functions of the vevs of the corresponding extra flavon fields that induce step-II (and III) breaking, and other physical scales involved (see Section~\ref{sec:model} for a specific realization).

\subsection{Lepton sector}
\label{subsec:lepton}

{\bf Step I:} For leptons, the presence of a Majorana mass term for neutrinos yields a different structure for the natural extrema of the flavon fields~\cite{Alonso:2013nca}
\begin{align}\label{eq:nat_min_l}
Y_e^{(0)}=\mathrm{diag}\left(0,0,y_\tau\right)\,,\quad
Y_\nu^{(0)}=
\begin{pmatrix}
i y_1 & 0 & 0\\
0 & \frac{i}{\sqrt{2}}y_2 & \frac{1}{\sqrt{2}}y_2\\
0 & \frac{i}{\sqrt{2}}y_3 & -\frac{1}{\sqrt{2}}y_3\\
\end{pmatrix}
\,,
\end{align}
that break the flavor group to the maximal subgroup
\begin{align}
\mathrm{SU(3)}_\ell\times\mathrm{SU(3)}_E\times\mathcal{O}(3)_{\nu_R}\to\mathrm{SU(2)}_E\times\mathrm{U(1)}_l\,,
\end{align}
where the $\mathrm{U(1)}_l$ subgroup is defined as
\begin{align}
\mathrm{U(1)}_l:\; \exp(i \alpha \lambda_3^{\prime\,\ell})\otimes\exp(i \frac{\sqrt{3}}{2} \alpha \lambda_8^E)\otimes\exp(i \alpha \lambda_7^{\nu_R})\,,
\end{align}
with $\lambda_3^\prime=\mathrm{diag}(0,1,-1)$. Again, the induced Yukawa pattern provides a good starting point for the description of the lepton masses and mixing angles.

\smallskip
{\bf Step II:} Perturbations will in turn induce additional breaking of the flavor group. Here we assume that the charged lepton {Yukawas} are generated at very high energies, i.e.
\begin{align}\label{eq:Yl_pert}
Y_e^{(1)}=\mathrm{diag}\left(y_e,y_\mu,0\right)\,,
\end{align}
with $|y_e|\ll|y_\mu|\ll |y_\tau|$, implying the following breaking pattern
\begin{align}
\mathrm{SU(2)}_E\times\mathrm{U(1)}_l\to\mathrm{U(1)}_{\mu-\tau}\,.
\end{align}
{Note that this symmetry breaking pattern requires the perturbations in the charged-lepton Yukawa to be flavor diagonal, as shown in Eq.~\ref{eq:Yl_pert}, therefore predicting no charged lepton flavor violation.}

{\bf Step III:} Finally, the $\mathrm{U(1)}_{\mu-\tau}$ symmetry given by
\begin{align}
\mathrm{U(1)}_{\mu-\tau}:\; \exp(i \alpha \lambda_3^{\prime\,\ell})\otimes\exp(i\alpha \lambda_3^{\prime\,E})\otimes\exp(i \alpha \lambda_7^{\nu_R})\,,
\end{align}
gets broken around the TeV scale by perturbations of the neutrino Yukawa.  As shown in Ref.~\cite{Alonso:2013nca}, these small perturbations allow to fully accommodate for the PMNS matrix.

\subsection{Couplings of the lightest \texorpdfstring{$Z'$}{Z'} bosons}
\label{subsec:couplings}

The breaking pattern illustrated above leads to two massive neutral gauge boson with masses around the TeV scale associated to the $\mathrm{U(1)}_q\times\mathrm{U(1)}_{\mu-\tau}$ flavor symmetry  while the other flavour gauge bosons are much heavier.

{\bf Quarkphilic $\hat Z_q$:} The fermion current associated to the $\hat Z_q$ field in the gauge basis ($\mathcal L \supset g_q \hat Z_{q\,\mu} \hat J_{Z_q}^\mu$) is
\begin{align}
\hat J_{Z_q}^\mu &= -\frac{1}{2}\,\overline{q^0_{kL}}\gamma_\mu q^0_{kL}\,   -\frac{1}{2}\,\overline{u_{kR}^0}\gamma_\mu u_{kR}^0\,   -\frac{1}{2}\,\overline{d_{kR}^0}\gamma_\mu d_{kR}^0\,   \nonumber\\
&\quad\,+\overline{q^0_{3L}}\gamma_\mu q^0_{3L}\,   + \,\overline{t_R^0}\gamma_\mu t_R^0\,   + \,\overline{b_R^0}\gamma_\mu b_R^0\,   \,, 
\end{align}
where $k=1,2$.  In order to diagonalize $Y_{u(d)}$ after the perturbation in Eq.~\eqref{eq:final_Yuk}, we introduce the unitary  matrices $U_{u\left( d \right)}^{L,R}$, such that $M_{u\left( d \right)}^{}\propto U_{u\left( d \right)}^{L\dag }Y_{u\left( d \right)}^{}U_{u\left( d \right)}^R$. To the leading order in $\delta_{u(d)}^i$ expansion we find
\begin{align}
U^L_{u(d)}\simeq
\begin{pmatrix}
1 & 0 & \frac{\delta^1_{u(d)}}{y_{t(b)}}\\[5pt]
0 & 1 & \frac{\delta^2_{u(d)}}{y_{t(b)}}\\[5pt]
-\frac{\delta^1_{u(d)}}{y_{t(b)}} & -\frac{\delta^2_{u(d)}}{y_{t(b)}} & 1
\end{pmatrix}
\,\mathcal{R}(\theta_{12}^{u(d)}),\quad U^R_{u(d)}\simeq \mathbb{1}\, ,
\end{align}
where $\theta_{12}^{u(d)}$ is an arbitrary $1-2$ rotation angle, determined beyond the leading order. Note that the CKM mixing matrix, $V_{\rm{CKM}}=U^{L\dagger}_{u}U^L_{d}$, can already be adjusted at this order with the appropriate choice of parameters. Finally, in the mass basis we get
\begin{align}
\begin{aligned}
\hat J_{Z_q}^\mu &= -\frac{1}{2}\,\overline{u_k}\gamma_\mu u_k -\frac{1}{2}\,\overline{d_k}\gamma_\mu d_k\, + \,\overline{t}\gamma_\mu t\,  + \,\overline{b}\gamma_\mu b\,  \\
&\quad\,+\left(\Gamma_{ij}^{u_L}\,\overline{u_{iL}}\gamma_\mu u_{jL}+\Gamma_{ij}^{d_L}\,\overline{d_{iL}}\gamma_\mu d_{jL}+{\rm h.c.}\right)\,,
\end{aligned}
\end{align}
with $i,j=1,2,3$ and the coupling matrices defined as
\begin{align}\label{eq:Gamma}
\Gamma^{u_L(d_L)}_{ij}=\frac{3}{2}\,\left[\left(U^{L}_{u(d)}\right)^*_{3i}\left(U^L_{u(d)}\right)_{3j}-\delta_{3i}\delta_{3j}\right]\,.
\end{align}
Note that the flavor universality of the gauge interactions with respect to the first two generations guarantees a suppression of the FCNCs among them, which are generated only at $\mathcal{O}(\delta^2)$. {An interesting scenario is given by the limit $U^L_{u}\to\mathbb{1}$}, where the coupling matrices read
\begin{align}
\begin{aligned}
\Gamma^{u_L}=0\,,\quad\Gamma^{d_L}=\frac{3}{2}
\begin{pmatrix}
\left|V_{td}\right|^2 & V_{ts}V_{td}^* & V_{tb}V_{td}^*\\
V_{ts}^*V_{td} & \left|V_{ts}\right|^2 & V_{tb}V_{ts}^*\\
V_{tb}^*V_{td} & V_{tb}^*V_{ts} & \left|V_{tb}\right|^2-1
\end{pmatrix}
\,.
\end{aligned}
\end{align}
{This limit, characterized by having no FCNCs in the up-quark sector, is {realized in the explicit model} in Section~\ref{sec:model}.} {In what follows, we will not consider the scenarios with FCNCs in the up-quark sector, given their strong model dependency.}

{\bf Leptophilic $\hat Z_\ell$:} Since we assume that the charged lepton {Yukawa} matrix is generated at a high scale where ${\rm U(1)}_{\mu-\tau}$ is unbroken, the $\hat Z_\ell$ charged lepton current in the mass basis is given by ($\mathcal L \supset g_\ell \hat Z_{\ell\,\mu} \hat J_{Z_\ell}^\mu$),
\begin{equation}
\hat J_{Z_\ell}^\mu =-\overline{\tau}\gamma_\mu \tau\, +\,\overline{\mu}\gamma_\mu \mu\,.
\end{equation}

{\bf Gauge-boson mixing:} At the scale where the two $\mathrm{U(1)}$'s are broken, the most general quadratic Lagrangian in the unitary gauge reads~\footnote{For simplicity, and without loss of generality, 
we assume that the new gauge bosons decouple from the SM gauge sector and there is no relevant 
kinetic or mass mixing with the SM gauge fields.}
\begin{align}
\begin{aligned}
\mathcal{L}&\subset-\frac{1}{4}\hat Z_{q\,\mu\nu}\hat Z_q^{\mu\nu}-\frac{1}{4}\hat Z_{\ell\,\mu\nu}\hat Z_\ell^{\mu\nu}-\frac{\sin\chi}{2}\hat Z_{q\,\mu\nu}\hat Z_\ell^{\mu\nu}\\
&\quad+\frac{1}{2} \hat M_{Z_q}^2\hat Z_{q\,\mu} \hat Z_q^\mu+\frac{1}{2} \hat M_{Z_\ell}^2\hat Z_{\ell\,\mu} \hat Z_\ell^\mu+\delta\hat M^2 \hat Z_{q\,\mu} \hat Z_\ell^\mu\,,
\end{aligned}
\end{align}
where $\chi$ and $\delta\hat M$ parametrize the kinetic and mass gauge mixing, respectively. The kinetic and mass mixing terms can be removed by means of a non-unitary and an orthogonal rotation, respectively (see Refs.~\cite{Babu:1997st,Langacker:2008yv} for more details). These are given by
\begin{align}\label{eq:Zpmass_rot}
\begin{pmatrix}
\hat Z_\ell\\
\hat Z_q
\end{pmatrix}
=
\begin{pmatrix}
1 & -t_\chi\\
0 & 1/c_\chi
\end{pmatrix}
\begin{pmatrix}
c_\xi & -s_\xi\\
s_\xi & c_\xi
\end{pmatrix}
\begin{pmatrix}
Z_1\\
Z_2
\end{pmatrix}
\,,
\end{align}
{where we used the following abbreviations: $c_\xi\equiv\cos\xi$, $s_\xi\equiv\sin\xi$, and $t_\xi\equiv\tan\xi$, and similarly for $\chi$. The} induced mass-mixing angle is defined as
\begin{align}
{t_{2\xi}}=\frac{-2 c_\chi\left(\delta\hat M^2- s_\chi \hat M_{Z_\ell}^2\right)}{\hat M_{Z_q}^2-c_{2\chi}\hat M_{Z_\ell}^2-2\,s_\chi\delta\hat M^2 }\,.
\end{align}
The gauge boson eigenstate masses are
{\small
\makeatletter
    \def\tagform@#1{\maketag@@@{\normalsize(#1)\@@italiccorr}}
\makeatother
\begin{equation}
M_{Z_{1,2}}^2 = \frac{\hat M_{Z_\ell}^2 + \hat M_{Z_q}^2-2\delta\hat M^2 s_\chi \pm\sqrt{\left(\hat M_{Z_\ell}^2 - \hat M_{Z_q}^2\right)^2+4\Delta}}{2 c^2_\chi} \,,
\end{equation}
}
where
\begin{align}
\Delta=\hat M_{Z_\ell}^2 \hat M_{Z_q}^2s_\chi^2+\delta\hat M^4-\left(\hat M_{Z_\ell}^2 + \hat M_{Z_q}^2\right)\delta\hat M^2 s_\chi\,.
\end{align}
Finally, the interactions of the gauge bosons with fermions in the mass eigenstate basis are given by
\begin{align}
\begin{aligned}
\mathcal{L}&\supset \left[\left(c_\xi-t_\chi s_\xi\right) g_\ell \hat J_{Z_\ell}^\mu+\frac{s_\xi}{c_\chi}\, g_q \hat J_{Z_q}^\mu\right]Z_{1\,\mu}\\
&+ \left[\frac{c_\xi}{c_\chi}\, g_q \hat J_{Z_q}^\mu -\left(s_\xi +t_\chi c_\xi\right) g_\ell \hat J_{Z_\ell}^\mu\right]Z_{2\,\mu}\,.
\end{aligned}
\end{align}

\section{Explicit model example}
\label{sec:model}

\begin{table}
\centering
\begin{tabular}{| c c c c c c c c |}
\hline  
 & $\mathrm{SU(2)}_Q$ & $\mathrm{SU(2)}_U$ & $\mathrm{SU(2)}_D$ & $\mathrm{U(1)}_q$ & $\mathrm{SU(3)}_c$ & $\mathrm{SU(2)}_L$ & $\mathrm{U(1)}_Y$ \\
\hline
\hline    
$q_{jL}$ & 2 & 1 & 1 & $-1/2$ & 3 & 2 & $1/6$ \\   
$u_{jR}$ & 1 & 2 & 1 & $-1/2$ & 3 & 1 & $2/3$ \\   
$d_{jR}$ & 1 & 1 & 2 & $-1/2$ & 3 & 1 & $-1/3$ \\   
$q_{3L}$ & 1 & 1 & 1 & 1 & 3 & 2 & $1/6$ \\   
$t_{R}$ & 1 & 1 & 1 & 1 & 3 & 1 & $2/3$ \\   
$b_{R}$ & 1 & 1 & 1 & 1 & 3 & 1 & $-1/3$ \\ \rowcolor{RGray}
$\mathcal{U}_{jL}$ & 1 & 2 & 1 & $-1/2$ & 3 & 1 & $2/3$ \\  \rowcolor{RGray}
$\mathcal{D}_{jL}$ & 1 & 1 & 2 & $-1/2$ & 3 & 1 & $-1/3$ \\ \rowcolor{RGray}
$\mathcal{U}_{jR}$ & 2 & 1 & 1 & $-1/2$ & 3 & 1 & $2/3$ \\ \rowcolor{RGray}
$\mathcal{D}_{jR}$ & 2 & 1 & 1 & $-1/2$ & 3 & 1 & $-1/3$ \\   
\hline
\hline
$H$ & 1 & 1 & 1 & 0 & 1 & 2 & 1/2 \\   \rowcolor{RGray}
$\phi_u$ & $\bar 2$ & 2 & 1 & 0 & 1 & 1 & 0 \\  \rowcolor{RGray}
$\phi_d$ & $\bar 2$ & 1 & 2 & 0 & 1 & 1 & 0 \\ \rowcolor{RGray}
$\phi_{\rm mix}$ & 1 & 1 & $2$ & $-3/2$ & 1 & 1 & 0 \\
\hline
\end{tabular}
\caption{ \small \sf  Particle content for the quark sector. Particles added to the SM are shown in a gray background.}
\label{tab:content_q}
\end{table}

In this section we present an explicit realization of the framework presented in the previous section. We assume that the maximal flavor group in the absence of Yukawas, $\mathcal{G}$, is a local symmetry of nature that gets broken by the natural minima in Eqs.~\eqref{eq:nat_min_q} and~\eqref{eq:nat_min_l} at a very high scale leading to an unbroken $\mathrm{SU(2)}_Q\times\mathrm{SU(2)}_U\times\mathrm{SU(2)}_D\times\mathrm{SU(2)}_E\times\mathrm{U(1)}_q\times\mathrm{U(1)}_l$. For simplicity, we will ignore in this section the generation of neutrino masses, that is not directly relevant to our phenomenological analysis.

Following a similar approach to that in Refs.~\cite{Grinstein:2010ve,Alonso:2016onw}, we introduce a minimal set of fermions to cancel the gauge anomalies, {i.e. those of the flavor and SM groups and mixed anomalies among them}, see Tables~\ref{tab:content_q} and~\ref{tab:content_l}. The most general renormalizable Lagrangian compatible with the symmetries and particle content of the model reads
\begin{align}
\mathcal{L}&=\mathcal{L}_{\rm kin}-V(\phi_u,\phi_d,\phi_{\rm mix},\phi_e,\phi_\nu,H) \nonumber\\
&\quad+(y_t\, \overline{q_{3L}} \widetilde H t_R+y_b\, \overline{q_{3L}} H b_R+y_\tau\, \overline{\ell_{3L}} H \tau_R \nonumber\\
&\quad+\lambda_u\,\overline{q_{iL}} \widetilde H \mathcal{U}_{iR}+\lambda_u^\prime\,\overline{\mathcal{U}_{iL}}(\phi_u)_{ij}\,\mathcal{U}_{jR}+M_u\,\overline{\mathcal{U}_{iL}}u_{iR} \nonumber\\
&\quad+\lambda_d\, \overline{q_{iL}} H \mathcal{D}_{iR}+\lambda_d^\prime\,\overline{\mathcal{D}_{iL}}(\phi_d)_{ij}\,\mathcal{D}_{jR}+M_d\,\overline{\mathcal{D}_{iL}}d_{iR}\nonumber\\
&\quad+\lambda_{\rm mix}\,\overline{\mathcal{D}_{iL}} (\phi_{\rm mix})_i\, b_R+\lambda_{e1}\,\overline{\ell_{1L}} H \mathcal{E}_{1R}\nonumber\\
&\quad+\lambda_{e2}\,\overline{\ell_{2L}} H \mathcal{E}_{2R}+\lambda_{e1}^\prime\,\overline{\mathcal{E}_{iL}}(\phi_e)_i\, \mathcal{E}_{1R}\nonumber\\
&\quad+\lambda_{e2}^\prime\,\overline{\mathcal{E}_{iL}}(\widetilde\phi_e)_i\,\mathcal{E}_{2R}+M_e\,\overline{\mathcal{E}_{iL}}e_{iR}+h.c.)\,,
\end{align}
with $i,j=1,2$ and where $M_{u,d,e}$ and $\lambda_{u,d}^{(\prime)}$ are universal parameters for the first two generations, and $\widetilde\phi_e=i\sigma_2\phi_e^*$ with $\sigma_2$ the Pauli matrix.
We assume that the scalar flavons $\phi_{u,d,e}$ take a vev at a high scale with $\langle \phi_f\rangle\gg M_f$ ($f=u,d,e$). This gives rise to the Yukawa couplings of the first and second generation SM fermions. Through a flavor transformation, we can make $\langle \phi_u\rangle$ diagonal and $\langle \phi_d\rangle\to \langle \phi_d\rangle\, V$, with $V$ a $2\times2$ unitary matrix that, at leading order, corresponds to the Cabbibo matrix. After step II symmetry breaking, the Yukawa couplings are given by
\begin{align}
Y_{u(d)}=
\begin{pmatrix}
y_{u(d)}^{(2)} & 0\\
0 & y_{t(b)}
\end{pmatrix}
\,,\quad
Y_e=
\begin{pmatrix}
y_e^{(2)} & 0\\
0 & y_\tau
\end{pmatrix}\,.
\end{align}
At leading order in the $M_f \langle \phi_f\rangle^{-1}$ expansion, the (light generation) Yukawa matrix elements read
\begin{align}
\begin{aligned}
&y_u^{(2)}=\frac{\lambda_u M_u}{\lambda_u^\prime} \langle \phi_u\rangle^{-1}\,,\quad y_d^{(2)}=\frac{\lambda_d M_d}{\lambda_d^\prime}V^\dagger \langle \phi_d\rangle^{-1}\,,\\
&(y_e^{(2)})_{ii}=\frac{\lambda_{ei} M_e}{\lambda_{ei}^\prime v_e}\,,
\end{aligned}
\end{align}
with $\langle \phi_e\rangle=(v_e\; 0)^\intercal$. On the other hand, the extra fermions acquire a mass proportional to the vevs of the flavon fields, which are assumed to be large, and they decouple at low energies. The vevs of the flavon fields break the flavor group down to $\mathrm{U(1)}_q\times\mathrm{U(1)}_{\mu-\tau}$. The unbroken $\mathrm{U(1)}_{\mu-\tau}$ symmetry ensures that the charged lepton Yukawa couplings are diagonal, and therefore there are no flavor violating couplings in the charged lepton sector.

Finally in the step III, the scalar flavons $\phi_{\rm mix}$ and $\phi_\nu$ develop a vev around the TeV scale giving a heavy mass to the neutral gauge bosons associated to the $\mathrm{U(1)}$'s ($\hat M_{Z_{q(\ell)}}\sim g_{q(\ell)}\langle\phi_{{\rm mix}(\nu)}\rangle$). Mixing among the third and the first two generations of quarks in the down-quark sector is generated,
\begin{align}
Y_d\stackrel{\langle\phi_{\rm mix}\rangle}{\longrightarrow}Y_d=
\begin{pmatrix}
y_d^{(2)} & y_d^{\rm mix}\\
0 & y_b
\end{pmatrix}
\,,
\end{align}
where, at leading order in $\langle\phi_d\rangle^{-1}\langle\phi_{\rm mix}\rangle$,
\begin{align}
y_d^{\rm mix}=y_d^{(2)}\,\frac{\lambda_{\rm mix}\langle\phi_{\rm mix}\rangle}{M_d}\,.
\end{align}
Note that in this model $V_{ub}\simeq(y_d^{\rm mix})_1/y_b$ and $V_{cb}\simeq(y_d^{\rm mix})_2/y_b$, and therefore in order to accommodate the measured values of the CKM matrix elements the hierarchy $\lambda_{\rm mix} \langle\phi_{\rm mix}\rangle \gtrsim M_d$ has to be enforced.

\begin{table}[t]
\centering
\begin{tabular}{| c c c c c c |}
\hline  
 & $\mathrm{SU(2)}_E$ & $\mathrm{U(1)}_l$ & $\mathrm{SU(3)}_c$ & $\mathrm{SU(2)}_L$ & $\mathrm{U(1)}_Y$ \\
\hline
\hline    
$\ell_{jL}$ & 1 & $\delta_{j2}$ & 1 & 2 & $-1/2$ \\   
$e_{jR}$ & 2 & $1/2$ & 1 & 1 & $-1$ \\  
$\ell_{3L}$ & 1 & $-1$ & 1 & 2 & $-1/2$ \\
$\tau_R$ & 1 & $-1$ & 1 & 1 & $-1$ \\  \rowcolor{RGray} 
$\mathcal{E}_{jL}$ & 2 & $1/2$ & 1 & 1 & $-1$ \\  \rowcolor{RGray}
$\mathcal{E}_{jR}$ & 1 & $\delta_{j2}$ & 1 & 1 & $-1$ \\  
\hline
\hline
$H$ & 1 & 0 & 1 & 2 & 1/2 \\   \rowcolor{RGray}
$\phi_e$ &  $2$ & $1/2$ & 1 & 1 & 0 \\  \rowcolor{RGray}
$\phi_\nu$ &  1 & $-1$ & 1 & 1 & 0 \\
\hline
\end{tabular}
\caption{ \small \sf  Particle content for the lepton sector. Particles added to the SM are shown in a gray background.}
\label{tab:content_l}
\end{table}

Finally, as we will discuss in Section~\ref{sec:Pheno} (see Eq.~\eqref{eq:combined_pheno}), a relatively large amount of mass mixing between the gauge bosons associated to the flavored $\mathrm{U(1)}$'s is needed in order to accommodate the $b\to s\ell^+\ell^-$ anomalies. In the minimal framework presented in this section, the only connection between the two sectors is given by the portal interaction $(\phi_{\rm mix}^\dagger\phi_{\rm mix})(\phi_\nu^\dagger\phi_\nu)$. This interaction is however unable to induce {the required mixing at sufficient level, which in the minimal model is absent even at the one-loop order. Therefore, the} minimal model presented here has to be extended. The necessary amount of mixing can be easily accounted for by the inclusion of additional particles, either fermions or scalars, charged under both $\mathrm{U(1)}_q$  and $\mathrm{U(1)}_{\mu-\tau}$, and with a mass around the TeV scale. For the purpose of this study it is not necessary to provide a precise realization of such extensions but we rather consider the gauge boson mixing as a free parameter.

\section{Phenomenological implications}
\label{sec:Pheno}

\subsection{Rare transitions: \texorpdfstring{$b \to s \mu^+ \mu^-$}{b->sll}}

We start with the effective Hamiltonian for $b \to s \mu^+ \mu^-$ transitions,
\begin{equation}
\mathcal{H}_{\textrm{eff}} = -\frac{4 G_F}{\sqrt{2}} V_{tb} V_{ts}^* \frac{e^2}{16\pi^2} \sum_i 
(C_i^{\mu\mu} O^{\mu\mu} _i + C^{\prime \mu\mu}_i O^{\prime \mu\mu}_i )+ \textrm{h.c.}
\end{equation}
In our model only a contribution to the operator
\begin{equation}
O^{\mu\mu}_9 \equiv (\bar s_L \gamma^\nu b_L)~ (\bar \mu \gamma_\nu \mu)\,,
\end{equation}
is generated. For our numerical analysis we rely on the results of the fit for the Wilson coefficients 
reported in Ref.~\cite{Descotes-Genon:2015uva} (see also Ref.~\cite{Altmannshofer:2015sma}), 
where the best fit is $C^{\mu\mu}_{9}|_{\textrm{NP}}=-1.09\pm0.22$ at the $1\sigma$ level. In our case, we have
\begin{align}\label{eq:C9}
C_9^{\mu\mu}|_{\textrm{NP}} &=\frac{g_q g_\ell \Gamma_{bs}^*}{V_{tb}V_{ts}^*}\frac{c_\xi s_\xi}{c_\chi}\left[\frac{1-t_\xi t_\chi}{M_{Z_1}^2}-\frac{1+ t_\chi/t_\xi}{M_{Z_2}^2}\right]\Lambda_\nu^2\,,
\end{align}
where $\Lambda_v =\sqrt{\sqrt{2} \pi/(G_F \alpha_{em})}=7$~TeV.

Naive {effective field theory} power counting suggests that the contribution due to pure kinetic mixing is expected to be {additionally} suppressed by a factor $\sim m_b^2 / M_{Z_{1,2}}^2$. {This is due to the presence of derivatives from the field-strength tensors.} Indeed, we explicitly checked that $C_9^{\mu\mu}|_{\textrm{NP}}$ in Eq.~\eqref{eq:C9} vanishes in the limit $\delta \hat M^2,m_b \to 0$. Therefore, from now on, we set $\chi$ to zero and allow for non-zero mass mixing $\delta \hat M^2 = \hat M_{Z_q} \hat M_{Z_\ell} \epsilon$, where the parameter $\epsilon$ is expected to be small. Expanding in $\epsilon$, we find
\begin{equation}\label{eq:c9}
C_9^{\mu\mu}|_{\textrm{NP}}
=-\frac{\Gamma_{bs}^*}{V_{tb}V_{ts}^*} \left(\frac{g_q \Lambda_v}{M_{Z_2}} \right) \left(\frac{g_\ell \Lambda_v}{M_{Z_1}} \right) \epsilon + \mathcal{O}(\epsilon^2)~.
\end{equation}

\subsection{\texorpdfstring{$\Delta F=2$}{DF=2} processes}

Flavor violating $Z'$ couplings to $b$ and $s$ unavoidably induce tree level contribution to $B_s - \bar B_s$ mixing. In our model
\begin{equation}
\Delta R_{B_{s}} = {\left|g_q \Gamma_{bs}\right|^2}\left(\frac{s_\xi^2}{M_{Z_1}^2}+\frac{c_\xi^2}{M_{Z_2}^2}\right) \left( \frac{g^2 (V_{tb} V_{ts}^*)^2}{16\pi^2 v^2} S_0\right)^{-1},
\end{equation}
where the SM loop factor $S_0\simeq2.3$. Expanding in $\epsilon$, we find
\begin{equation}
\Delta R_{B_{s}} = \left|\frac{\Gamma_{bs}}{V_{tb} V_{ts}^{*}}\right|^2 \left(\frac{g_q \Lambda_v}{M_{Z_2}} \right)^2 \left(\frac{2 s_W^2}{S_0} \right)+\mathcal{O}(\epsilon^2)\,,
\end{equation}
where $s_W$ is the sine of Weinberg angle. Requiring NP contributions to the mixing amplitude to be at most $\mathcal{O}(10\%)$,
we find the following condition
\begin{equation} \label{eq:bsmix}
	\left|\frac{\Gamma_{bs}}{V_{tb} V_{ts}^{*}} \frac{g_q \Lambda_v}{M_{Z_2}}\right|\lesssim 0.7~. 
\end{equation}
In the numerical fit we take NP contributions to the mixing amplitude to be $\Delta R_{B_{s}}=-0.10\pm0.07$ (see discussion in Ref.~\cite{Buttazzo:2016kid}).

\begin{figure}[ht]
\includegraphics[width=.85\hsize]{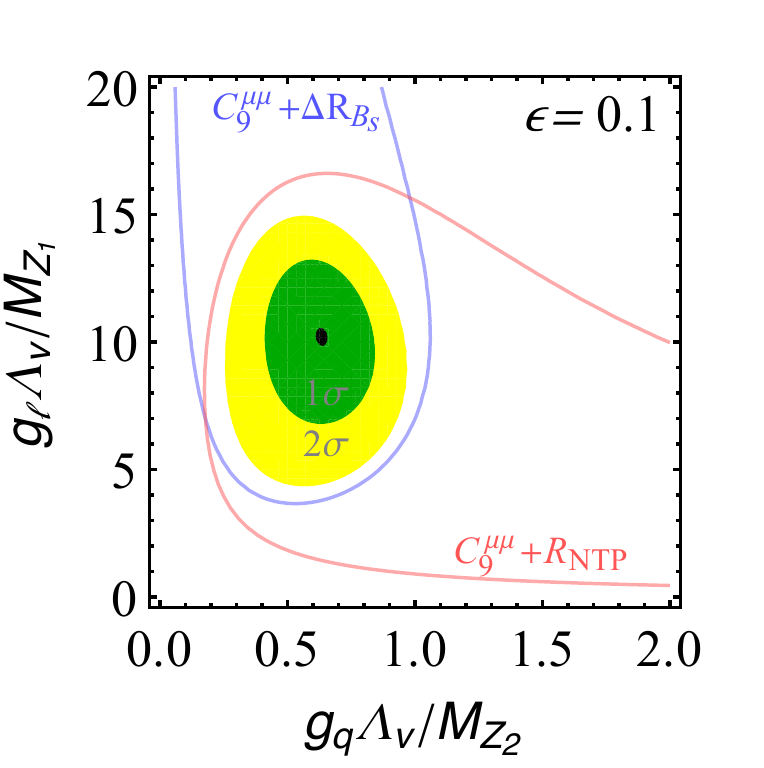}
\caption{\small \sf Combined fit to $B_s$ mixing, neutrino trident production and $b \to s \mu^+ \mu^-$ observables assuming $\epsilon=0.1$ (green - 1$\sigma$, yellow - 2$\sigma$). Relaxing the first (second) constraint is illustrated with red (blue). 
\label{fig:fit} }
\end{figure}

It is worth noting that the contribution to $\Delta F=2$ processes have necessarily constructive interference and are MFV-like (i.e.~a similar relative correction compared to the SM is also expected in $B_d-\overline{B}_d$ and Kaon mixing).  This conclusion would not hold if the  $\sigma_{d}^i$
terms in Eq.~(\ref{eq:final_Yuk}) were not negligible.  Tuning the $\sigma_{d}^i$ one can generate arbitrary contributions 
to  $\Delta F=2$ amplitudes  and relax the bound in Eq.~(\ref{eq:bsmix}). However, since this tuning does not find a natural explanation 
within our framework, we will not consider this possibility any further. Flavon fields also give tree-level MFV-like contributions to $\Delta F=2$ transitions. However, since they are SM singlets, their contributions to these processes receive an extra suppression of $\mathcal{O}\left(m_b^2/\langle \phi_{\rm{mix}}\rangle^2\right)$, and can be neglected. We checked this explicitly in the model example and also verified that flavon-Higgs box contributions can be neglected since they are not parametrically enhanced by large masses compared to the $Z^\prime$ contribution.

\subsection{Neutrino Trident Production}

Bounds on flavor-diagonal $Z^\prime$ couplings to muons can also arise from neutrino trident production (NTP), where a muon pair is created by scattering a muon-neutrino with a nucleon: $\nu_\mu N \to \nu N \mu^+\mu^-$~\cite{Altmannshofer:2014pba}. Note that as the flavor of the neutrino in the final state is not detected, one must sum over all three generations in the case of flavor-violating interactions. We obtain for the cross section of NTP
\begin{align}
\frac{\sigma_{\rm NP}}{\sigma _{\rm SM}} = \frac{1+\left(1+4s_W^2+2v^2V^{\rm{NP}}\right)^2}{1+\left(1+4s_W^2\right)^2}\,,
\end{align}
with
\begin{align}
V^{\rm{NP}}=g_\ell^2\left[{\left(\frac{c_\xi}{M_{Z_1}}\right)^2+\left(\frac{s_\xi}{M_{Z_2}}\right)^2}\right]~,
\end{align}
while 
expanding in $\epsilon$
\begin{equation}
V^{\rm{NP}}=\frac{g_\ell^2}{M_{Z_1}^2}+\mathcal{O}(\epsilon^2)~.
\end{equation}
The bound from the CCFR collaboration~\cite{Mishra:1991bv} is given by
\begin{align}
R_{\rm NTP}\equiv\sigma_{\rm exp}/\sigma_{\rm SM}=0.82\pm 0.28\,.
\end{align}
Requiring this constraint to be satisfied at the $2\sigma$ level implies
\begin{equation}\label{eq:NTP}
	\left|\frac{g_\ell \Lambda_v}{M_{Z_1}} \right| \lesssim 12~.
\end{equation}

\subsection{Combined fit to low-energy data}

Using the limits in Eq.~\eqref{eq:bsmix} and Eq.~\eqref{eq:NTP}, and plugging in Eq.~\eqref{eq:c9}, 
we find that the NP contribution to $C_9^{\mu\mu}$ can be 
\begin{equation}\label{eq:combined_pheno}
C_9^{\mu\mu}|_{\textrm{NP}} \simeq -\, 8 \times \epsilon~.
\end{equation}
We then conclude that  $\epsilon \sim \mathcal{O}(0.1)$ is required to reconcile low-energy 
constraints with the correct size of $C_9^{\mu\mu}$. 
To make this point more precise, we perform a combined fit to the set of measurements discussed above. The total likelihood as a function of three parameters: $g_q \Lambda_v/M_{Z_2}$, $g_\ell \Lambda_v/M_{Z_1}$ and $\epsilon$, is constructed by adding the corresponding $\chi^2$ terms from the three measurements: $C_9^{\mu\mu}|_{\textrm{NP}}$, $\Delta R_{B_{s}}$ and $R_{\rm NTP}$.  The preferred region for  
$g_q \Lambda_v/M_{Z_2}$ and $g_\ell \Lambda_v/M_{Z_1}$  at $1\sigma$ (green) and $2\sigma$ (yellow), setting 
$\epsilon = 0.1$,  is shown in Fig.~\ref{fig:fit}. {In the numerical analysis, we set $\Gamma_{b s} =V_{t b} V_{t s}^*$ as suggested by Eq.~\eqref{eq:Gamma}. Possible $\mathcal{O}(1)$ modifications of this parameter shift accordingly the preferred value of $g_q \Lambda_v/M_{Z_2}$.}

\begin{figure}[t]
\includegraphics[width=.9\hsize]{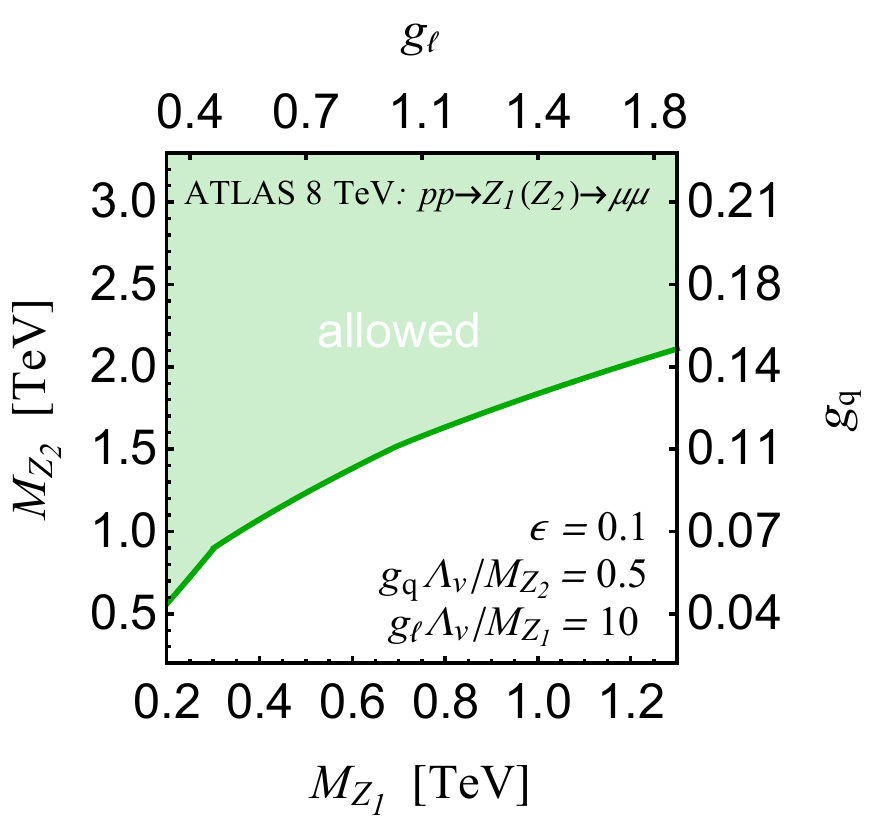}
\caption{\small \sf Allowed region in the $(M_{Z_1}, M_{Z_2})$ plane from direct searches for dimuon resonances in $pp \to Z_1 (Z_2) \to \mu^+ \mu^-$ at LHC~\cite{Aad:2014cka}. Here we fix $\epsilon = 0.1$, $g_q \Lambda_v / M_{Z_2}=0.5$, and $g_\ell \Lambda_v / M_{Z_1}=10$ in agreement with the preferred region from the low-energy fit. The mixing angle $\xi$ is small for these points, at most $\sim 0.1$.}
\label{fig:muon-limits}
\end{figure}

\subsection{Direct searches at LHC}

\begin{figure*}[t]
  \centering
  \begin{tabular}{ccc}
   \includegraphics[width=0.325\textwidth]{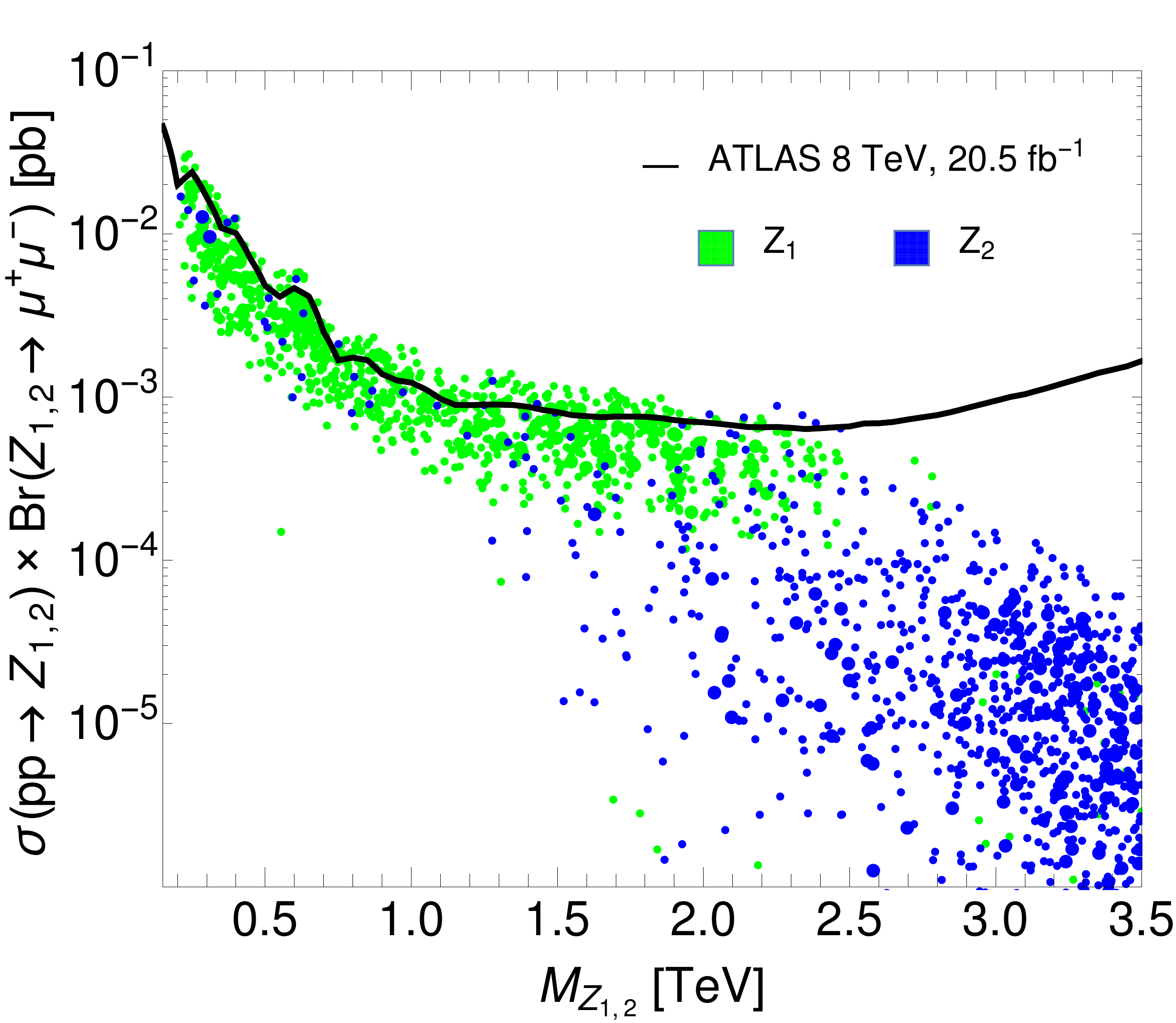} &
   \includegraphics[width=0.325\textwidth]{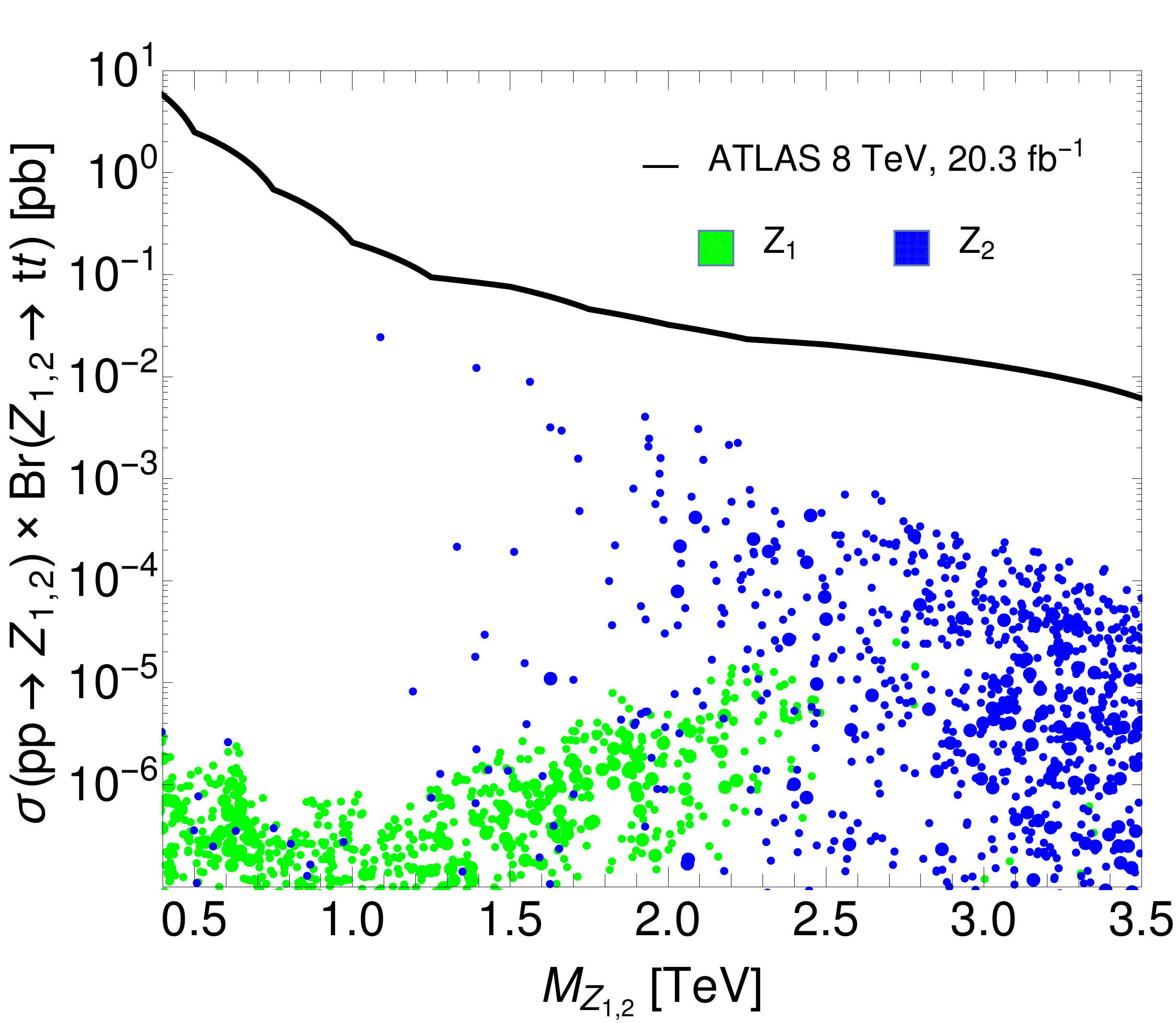} & 
   \includegraphics[width=0.325\textwidth]{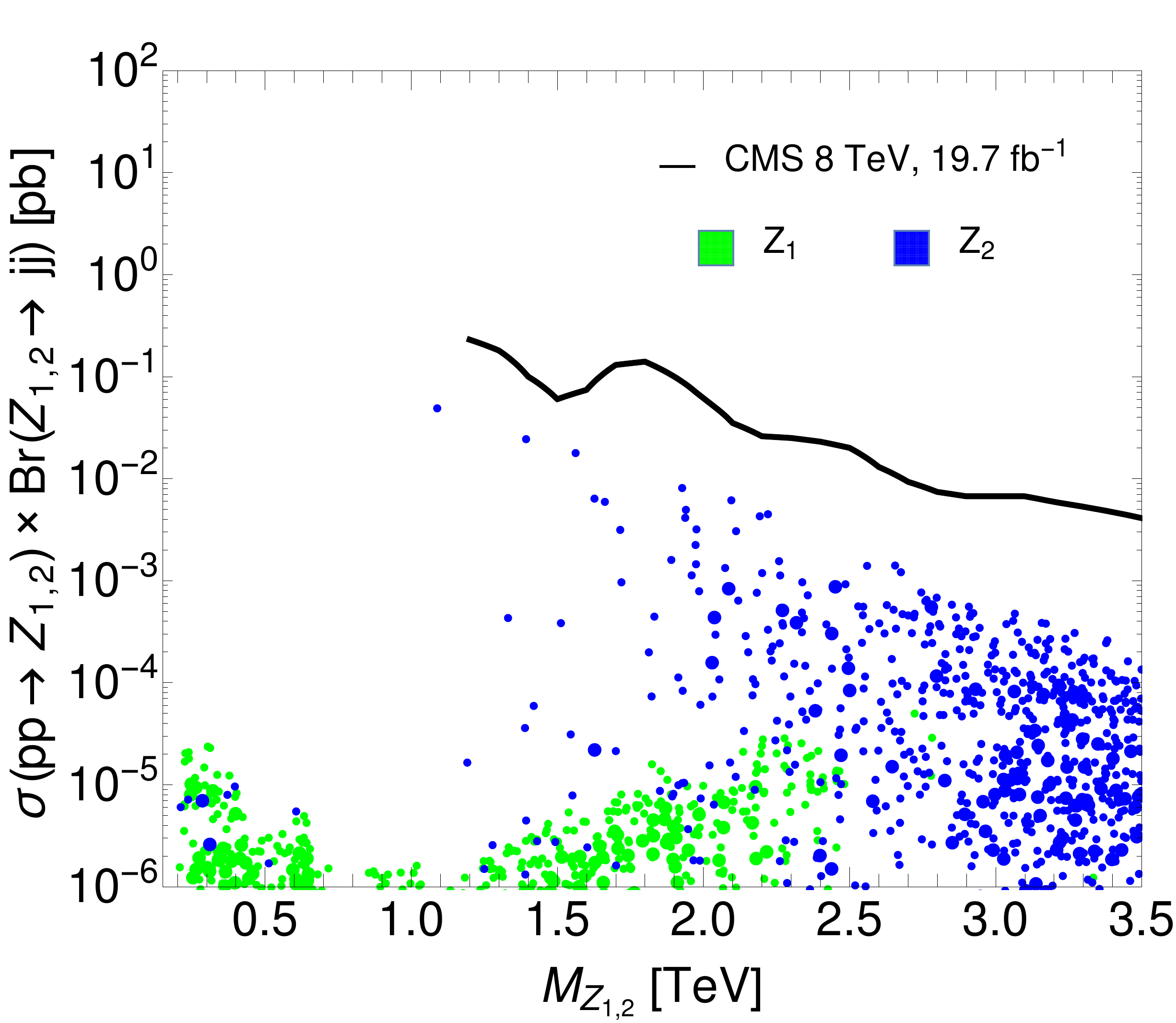}  \\
    \includegraphics[width=0.325\textwidth]{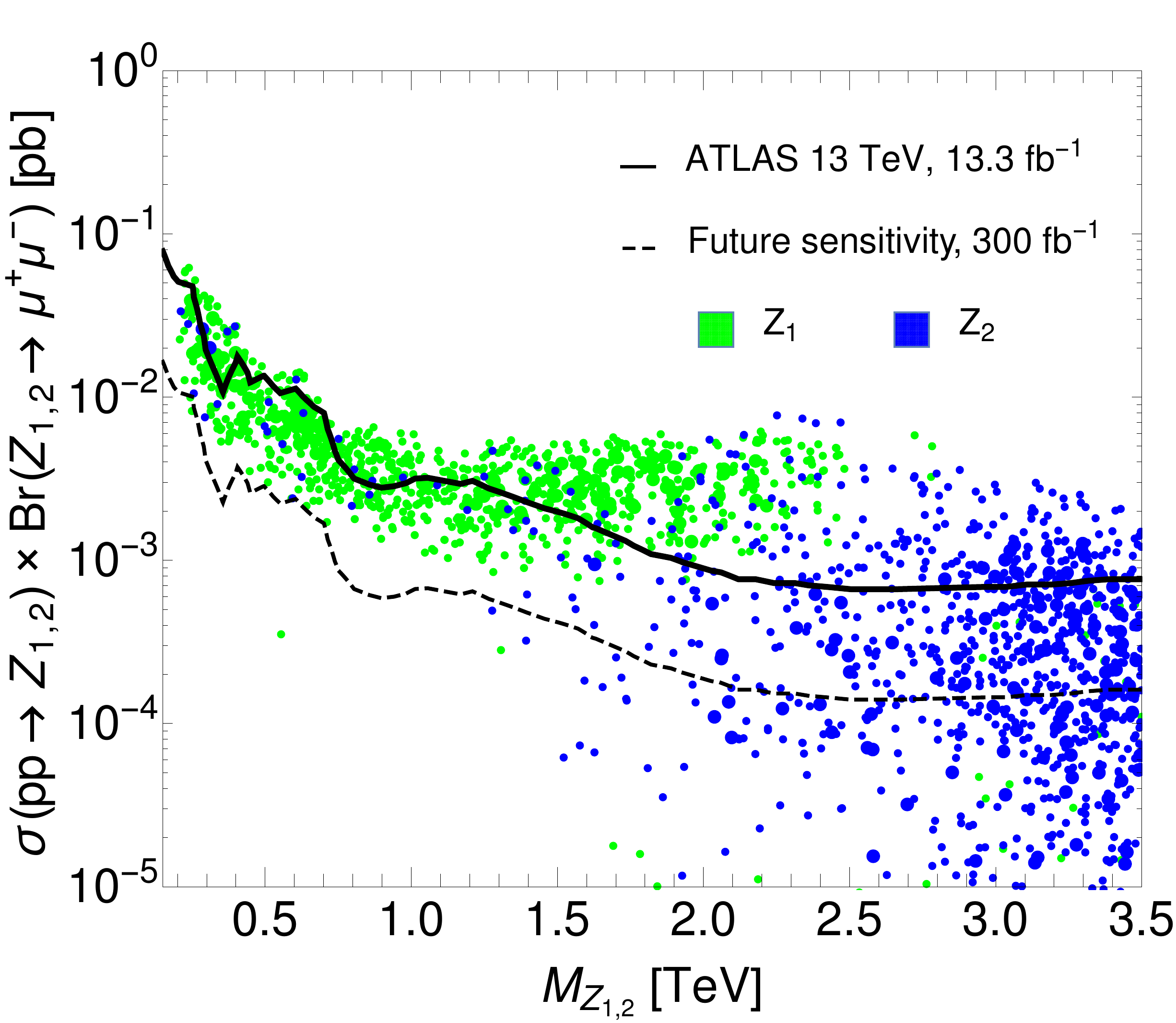} &
    \includegraphics[width=0.325\textwidth]{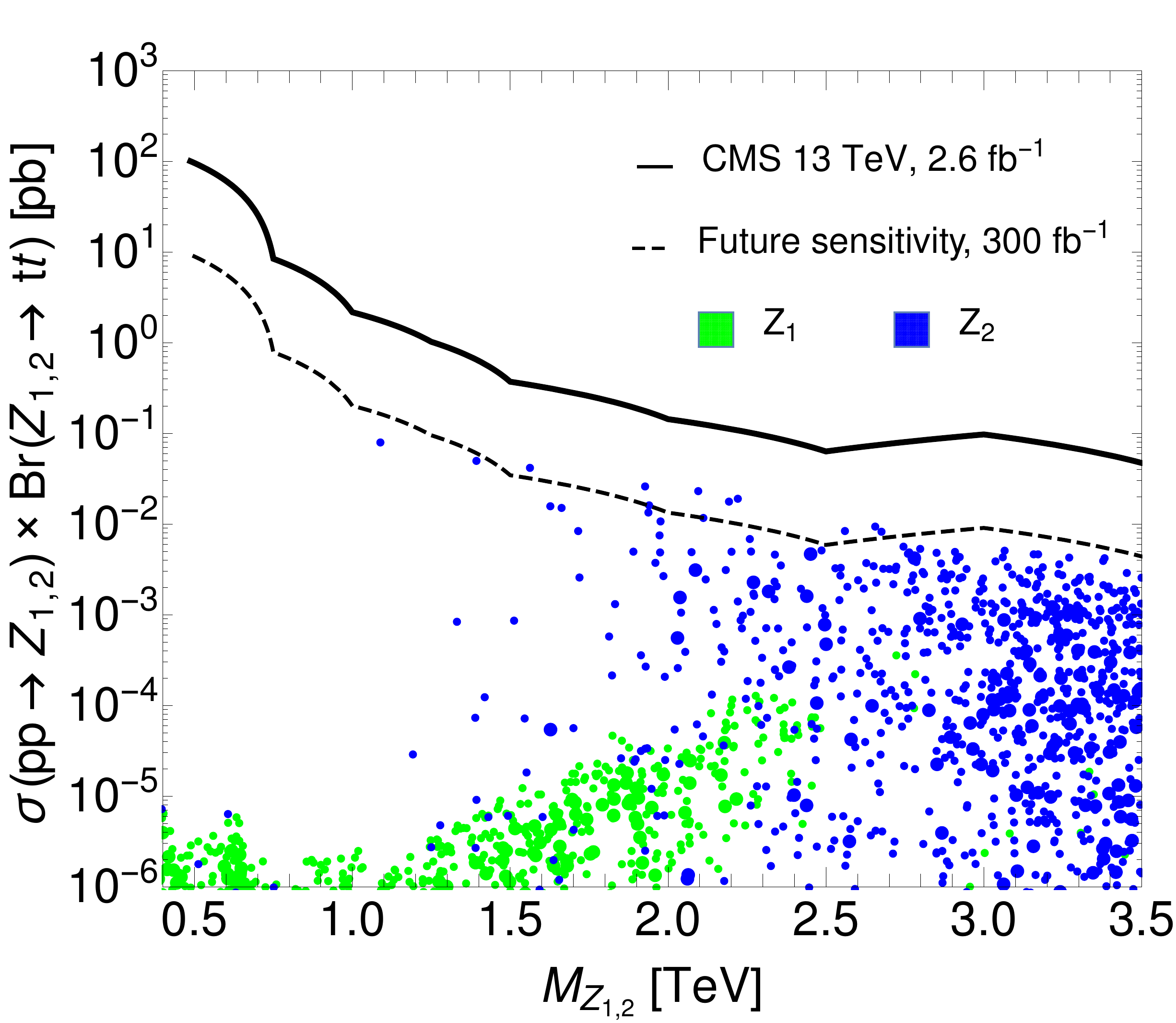} &
    \includegraphics[width=0.325\textwidth]{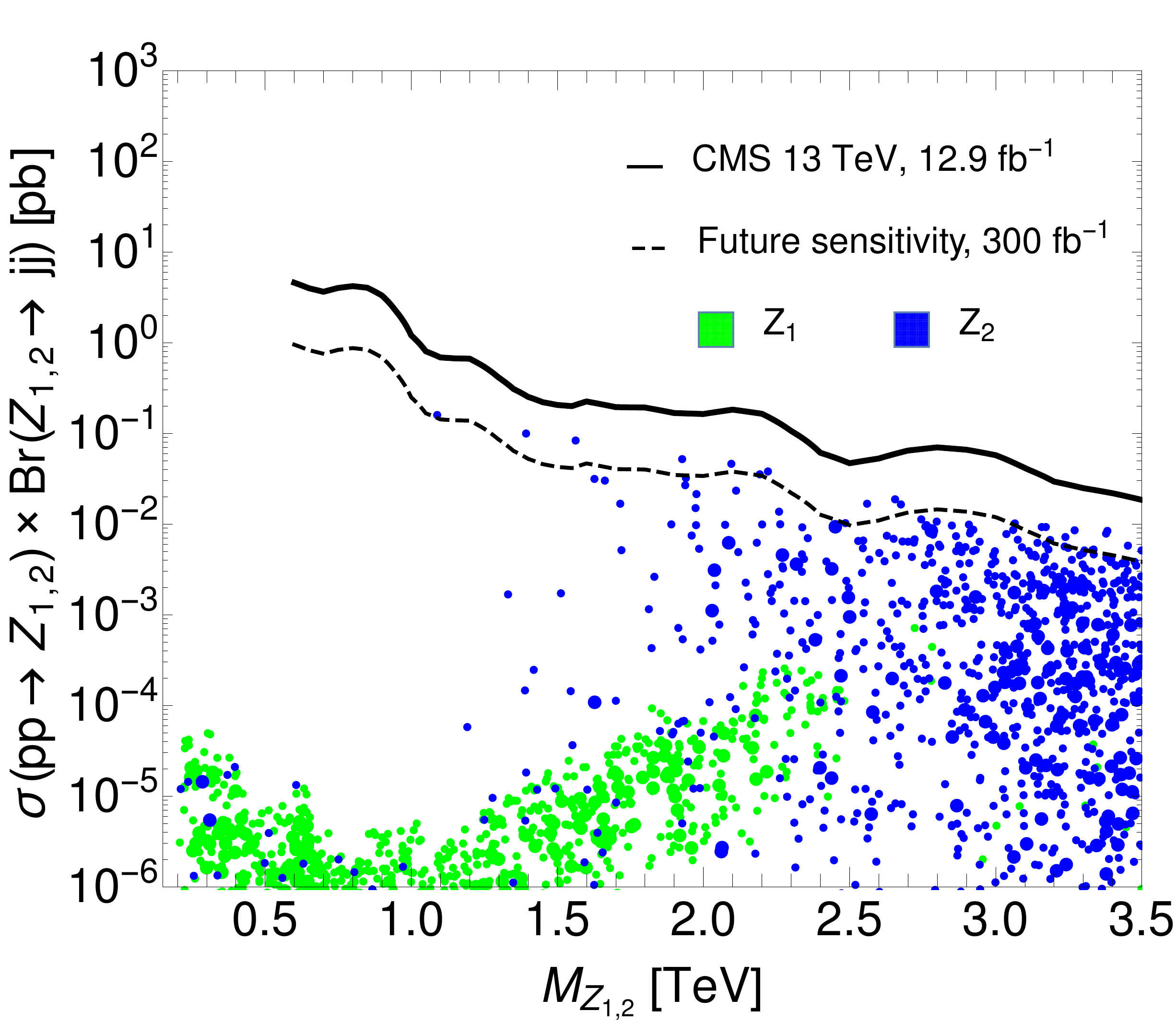}\\
    \end{tabular}
    \caption{ \small \sf Predictions for the LHC signals ($\sigma\times\mathcal{B}$) at 8 TeV (first row) and 13 TeV (second row) for $\mu^+ \mu^-$, $t\bar t$ and $jj$ resonance searches for $Z_1$ (green) and $Z_2$ (blue). Present limits are shown with black line, and future-projected with dashed line.
    \label{fig:LHC-pred}}
    \end{figure*}

In this subsection, we discuss the LHC phenomenology of the two $Z'$ bosons. In the limit of zero mass and kinetic mixing and assuming that right-handed neutrinos are heavy, {$Z_1 \simeq \hat Z_{\ell}$ (see Eq.~\eqref{eq:Zpmass_rot})} decays predominantly to $\tau^+ \tau^-$, $\mu^+ \mu^-$ and $\bar \nu \nu$ with the partial decay widths
\begin{equation}
\Gamma(\hat Z_{\ell} \to \mu \mu) = \Gamma(\hat Z_{\ell} \to \tau \tau) =  \Gamma(\hat Z_{\ell} \to \bar \nu \nu) = \frac{g_\ell^2}{12 \pi} M_{\hat Z_{\ell}}~,
\end{equation}
while ${Z_2 \simeq \hat Z_{q}}$ decays to $\bar t t$, $\bar b b$ and light jets with the partial widths
\begin{equation}
\Gamma(\hat Z_{q} \to b \bar b) = \Gamma(\hat Z_{\ell} \to j j) = \frac{g_q^2}{4 \pi} M_{\hat Z_{q}}~,
\end{equation}
where $jj=u\bar u+d \bar d + s\bar s +c \bar c$ and 
\begin{equation}
\Gamma(\hat Z_{q} \to t \bar t) = \Gamma(\hat Z_{q} \to b \bar b) \left(1+\frac{2m_t^2}{M_{\hat Z_{q}}^2}\right)\sqrt{1-\frac{4 m_t^2}{M_{\hat Z_{q}}^2}}~,
\end{equation}
in agreement with the general decay formula in Eq.~(B9) of Ref.~\cite{Bishara:2015mha}.
In the zero mixing limit, only $\hat Z_{q}$ is produced at the LHC. The total hadronic cross section in the narrow width approximation is given by
\begin{equation}
\sigma(q \bar q \to \hat Z_{q} ) = \frac{8 \pi^2}{3 M_{\hat Z_{q}} s_0}~ \Gamma(\hat Z_{q} \to q \bar q)~\mathcal{L}_{q \bar q}~,
\end{equation}
where $\mathcal{L}_{q \bar q}$ is the corresponding parton luminosity function. The dominant contribution is for $q = u,  d $. We use NNLO MMHT2014 PDF~\cite{Harland-Lang:2014zoa} set for numerical studies. Furthermore, we cross-checked the results using MadGraph~\cite{Alwall:2014hca}.

ATLAS searched for a narrow $Z'$ resonance in $pp$ collisions at 8 TeV decaying to $\mu^+ \mu^-$~\cite{Aad:2014cka}. The reported limits on $\sigma \times \mathcal{B}$ in the mass range [$150-3500$]~GeV  can be used to set constraints on the parameter space of our model. Production and decay formulas are easily generalized in the case of arbitrary mass mixing $\xi$. We compute the signal strength for both $Z'$ and confront with these limits.

Interestingly, we find that it is possible to have relatively light vectors ($\lesssim$~TeV) with $\mathcal{O}(1)$ couplings. In particular, fixing $\epsilon = 0.1$, $g_q \Lambda_v / M_{Z_2}=0.5$, and $g_\ell \Lambda_v / M_{Z_1}=10$, in agreement with the preferred region from the low-energy fit, we have performed a scan in the $(M_{Z_1}, M_{Z_2})$ plane. In Fig.~\ref{fig:muon-limits} we show in green the region allowed by present dimuon searches from Ref.~\cite{Aad:2014cka}. The most plausible scenario is the one with a mass hierarchy between the vectors, $M_{Z_1} < M_{Z_2}$, and with small mass mixing. That is, the lighter vector $Z_{1}$ is predominantly $\hat Z_{\ell}$, while the heavier $Z_{2}$ is predominantly $\hat Z_{q}$. In conclusion, requiring small (perturbative) couplings ($g_\ell\sim 2$ implies $\Gamma_{Z_1}/M_{Z_1}\sim0.3$), low-energy flavor data together with dimuon resonance searches require a relatively light leptophilic $Z'$, that might be probed in the near future.

In order to check the robustness of the above statement, we have performed an exhaustive parameter scan of the model randomly varying five input parameters with flat priors: $\hat M_{Z_q}$ and $\hat M_{Z_\ell}$ in the range [150,~4000]~GeV, $\epsilon$ and $g_q$ in the range [0,1], and $g_\ell$ in the range [0,2]. We  have constructed the combined likelihood function for the low-energy data, together with the dimuon search~\cite{Aad:2014cka} for both $Z'$s, in terms of the model parameters using the complete formulas (not expanded in small mixing). The best fit point gives $\chi^2_{\rm{min}}\approx 7$, while the SM point has $\chi^2_{\rm{SM}}\approx 27$. We have kept the points that provide a good fit to all data, namely $\Delta \chi^2\equiv \chi^2-\chi^2_{\rm{min}}  \lesssim 6$.

Shown in Fig.~\ref{fig:LHC-pred} are the corresponding predictions for $\sigma\times\mathcal{B}$ for $\mu^+ \mu^-$, $t\bar t$ and $j j$ resonance searches at LHC. In addition, we show the present limits from Refs.~\cite{ATLAS:2016cyf,Aad:2015fna,Khachatryan:2015sja,CMS:2016wpz,CMS:2016zte}, and estimate the future sensitivity with $300$~fb$^{-1}$. Interestingly enough, the unpublished $13$~TeV dimuon resonance search~\cite{ATLAS:2016cyf} is already probing the relevant region, with conclusive answers expected in the near-future data. On the other hand, we find the impact of the present (and future) $\bar t t$ and $j j$ searches to be less relevant.  We restricted our scan to the mass range of dimuon resonance searches reported by ATLAS and CMS (namely $M_{\mu\mu} \geq 150$ GeV) but it is interesting to note that a very light (almost leptophilic) $Z_1$ could evade the experimental bounds provided its gauge coupling is sufficiently small. In this case, $Z\to 4\mu$ provides a bound of $M_{Z_1}\gtrsim 30$ GeV~\cite{Altmannshofer:2014cfa,Altmannshofer:2014pba,delAguila:2014soa}, when the three-body decay $Z\to \mu^+\mu^-Z^\prime(\to\mu^+\mu^-)$ is kinematically open. {Finally, one can establish a rough estimate on the future mass reach at LHC using the ColliderReach tool for extrapolations~\cite{ColliderReach}. We find that by Run 3, with $300\,\textrm{fb}^{-1}$, this value should rise to about $5$~TeV.}

\section{Summary and conclusions}
\label{sec:conclusions}
The assumption of dynamically generated Yukawa couplings provides a natural explanation to the observed pattern of fermion masses and mixing angles, 
both in the quark and lepton sectors~\cite{Alonso:2013nca}. In this framework the maximal $[\mathrm{SU(3)}]^5\times \mathcal{O}(3)$ flavor group is assumed to be a local symmetry of nature, broken spontaneously (and in several steps) by flavon fields. 
In this paper we have shown that in this context, under reasonable assumptions about  the flavor symmetry breaking pattern,  it is possible to 
obtain an explanation of  the anomalies observed in
$b\to s\mu^+\mu^-$ transitions. The main features of the proposed model can be summarized as follows.
\begin{itemize}
\item Two $Z^\prime$ bosons arise as the lowest-lying resonances resulting from the gauging of the flavor group, one corresponding to a gauged $\mathrm{U(1)}_q$ and the other one to a gauged $\mu-\tau$ flavor symmetry. A small but non-vanishing mass-mixing among the 
 $Z^\prime$ bosons is required in order to accommodate the flavor anomalies.

\item  The flavor symmetry  acting on the light quark families ensures a partial protection for quark FCNCs,
which turn out to be sufficiently small to avoid the tight existing constraints while allowing for sizable effects in $b\to s\mu^+\mu^-$.
The model predicts no FCNCs in the charged lepton sector.

 \item Concerning $b\to s\mu^+\mu^-$ transitions, our model predicts $\left.C_9^{\mu\mu}\right|_{\rm NP}$ only, and maximal $\mu-e$ universality violation. Therefore, no deviations from the SM predictions in $B_s\to \mu^+\mu^-$ are expected but sizable effects in angular observables measuring lepton flavor universality violation {\cite{Descotes-Genon:2016hem,Capdevila:2016ivx,Serra:2016ivr}} should occur.
\end{itemize}

Present data already provides important restrictions on the parameter space of the model: for small gauge mixing, the bound from $B_s$ mixing implies $g_q/M_{Z_2}\lesssim0.1\;\mathrm{TeV}^{-1}$ while the constraint from NTP gives the bound $g_\ell/M_{Z_1}\lesssim 1.7 \;\mathrm{TeV}^{-1}$. Direct searches at LHC allow for a very light (almost leptophilic) $Z^\prime$ together with a heavier one (mostly quarkphilic) in the TeV domain. This full region of the parameter space will be explored in the near future by dimuon searches.

\bigskip
\acknowledgments{We thank our colleagues at the Physik-Institut for useful discussions. J.F. thanks the University of Zurich for hospitality during the completion of this work. The work of a A.C. is supported by a Marie Curie Intra-European Fellowship of the European Community's 7th Framework Programme under contract number PIEF-GA-2012-326948 and by an Ambizione Grant of the Swiss National Science Foundation. The work of J.F. is supported in part by the Spanish Government, by Generalitat Valenciana and by ERDF funds from the EU Commission [grants FPA2011-23778,FPA2014-53631-C2-1-P, PROMETEOII/2013/007, SEV-2014-0398]. J.F. also acknowledges VLC-CAMPUS for an ``Atracci\'o de Talent'' scholarship. A.G. and G.I. are supported by the Swiss National Science Foundation (SNF) under contract 200021-159720.}

\end{document}